\documentclass[conference]{IEEEtran}
\IEEEoverridecommandlockouts
\usepackage{cite}
\usepackage{amsmath,amssymb,amsfonts}
\usepackage{algorithmic}
\usepackage{graphicx}
\usepackage{textcomp}
\usepackage[dvipsnames]{xcolor}
\usepackage{subfig}
\usepackage{dcolumn}
\usepackage{bm}
\usepackage{physics}
\usepackage{float}
\usepackage{hyperref}
\def\BibTeX{{\rm B\kern-.05em{\sc i\kern-.025em b}\kern-.08em
    T\kern-.1667em\lower.7ex\hbox{E}\kern-.125emX}}
    
\begin{document}

\title{Flux pumping of Cooper pairs through a Josephson Energy-Suppression Pump}

\author{
\IEEEauthorblockN{1\textsuperscript{st} Angelo Greco}
\IEEEauthorblockA{
\textit{Politecnico di Torino (PoliTo)} \\
\textit{Istituto Nazionale di Ricerca Metrologica (INRiM)}\\
Torino, Italy \\
a.greco@inrim.it}
\and
\IEEEauthorblockN{2\textsuperscript{nd} Luca Fasolo}
\IEEEauthorblockA{
\textit{Politecnico di Torino (PoliTo)} \\
\textit{Istituto Nazionale di Ricerca Metrologica (INRiM)}\\
Torino, Italy \\
luca\_fasolo@polito.it}
\and
\IEEEauthorblockN{3\textsuperscript{rd} Vito Marino}
\IEEEauthorblockA{
\textit{Politecnico di Torino (PoliTo)} \\
Torino, Italy \\
vimaro18@gmail.com}
\and
\IEEEauthorblockN{4\textsuperscript{th} Emanuele Enrico}
\IEEEauthorblockA{
\textit{Istituto Nazionale di Ricerca Metrologica (INRiM)} \\
Torino, Italy \\
\textit{Trento   Institute for  Fundamental  Physics  and  Applications (INFN),  Trento,  Italy}\\
e.enrico@inrim.it}
}

\maketitle

\begin{abstract}
In this paper, we propose a novel kind of Josephson Energy Suppression Pump (JESP) controlled by a fully magnetic flux drive. The device presented here is composed of two superconducting loops interrupted at one side by superconducting nanowires which are joined together by a superconducting island. The phase difference developed at the edges of the nanowires by means of the magnetic flux threading the loops can collapse their Cooper condensates, leading to complete suppression of the Josephson energies. This mechanism allows to greatly reduce the leakage current when performing Cooper pair pumping by a pure magnetic pumping cycle without involving any gate modulation. The pumping capability of the JESP is studied through a master equation approach in the non-adiabatic case.
\end{abstract}


\section{\label{sec:introduction}Introduction}

\begin{figure}
\includegraphics[width=0.5\textwidth]{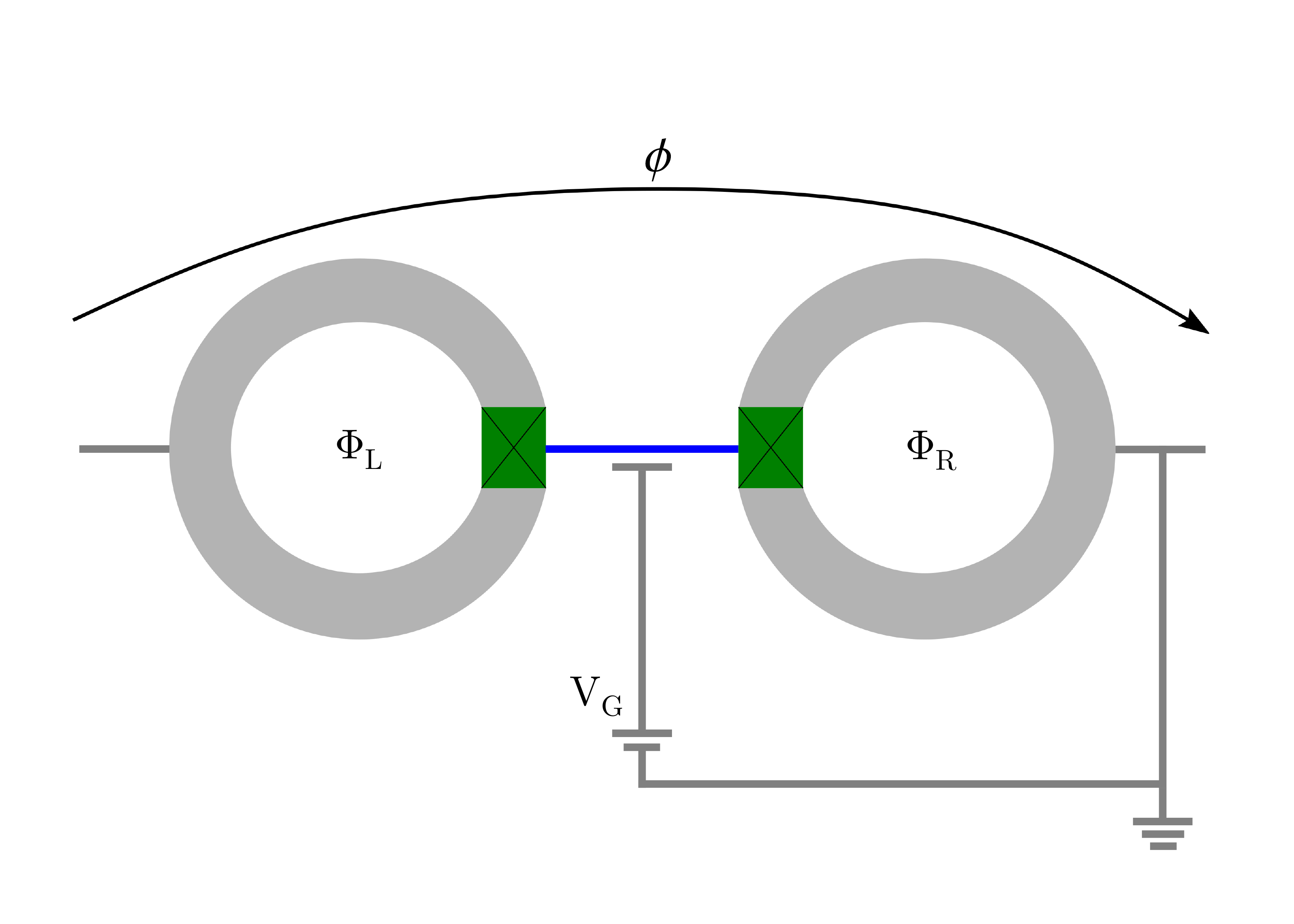}
\caption{\label{fig:sketch} Sketch of the structure of a JEST: $\Phi_\mathrm{L}$ and $\Phi_\mathrm{R}$ are the magnetic fluxes threading two superconducting loops (light grey). The (blue) island connects the two nanowires (green) and is set at the working point by the gate voltage $V_\mathrm{G}$. Two Josephson junctions are formed by the nanowires plus the superconducting island. A bias phase difference $\phi$ is kept constant at the two sides of the device.}
\end{figure}

Cooper-pair pumps have gained general recognition as an implementation of a driven quantum two-level system \cite{PhysRevB.84.235140} and represent good candidates to probe decoherence effects in presence of a dissipative environment\cite{PhysRevLett.105.030401, PhysRevB.82.134517,PhysRevB.83.214508}. Their operation is commonly based on the periodic modulation of some of their parameters, leading to a parametric cycle that guides Cooper-pair tunnelling through two or more quantized-charge-sensitive tunnel junctions. The intrinsic macroscopic coherence of this fully superconducting charge pumps, where the Cooper-pair condensate order parameter plays a key role, opens up the possibility to exploit the driven modulation of the Josephson energy in the framework of the so-called “quantum pumping” \cite{PhysRevB.58.R10135, Switkes1905, Giazotto2011}.  
In this view, the Cooper-pair sluice first proposed driving cycle\cite{Niskanen2003} intermixes magnetic and electric fields, ideally resulting in a negligible dynamical supercurrent leakage. Among the different pumping schemes investigated, the so-called flux-pumping promises a purely coherent and arbitrarily large pumped charge\cite{Gasparinetti2012}. There the Cooper pair sluice is driven by a constant gate voltage and the opening times of the two Superconducting Quantum Interference Devices (SQUIDs) based electrodes acting as valves are largely overlapping.  
The geometric properties of the parametric cycle reflect on the pumped charge in the adiabatic limit with an explicit relation that connects this latter to the geometric (Berry) phase. Experimental measurements of the accumulated Berry phase during each adiabatic pumping cycle have been performed, meanwhile, the dynamic and geometric current contributions have been separately identified\cite{Mottonen2008}. Landau-Zener transitions between energy levels limit the adiabatic regime and can be exploited to realize Landau-Zener-St\"uckelberg interferometers\cite{SHEVCHENKO20101} when coupled to geometric phases\cite{PhysRevLett.107.207002}. Such transitions introduce decoherence in the system and can be induced by charge noise or non-adiabatic drive.
The first experimental demonstration of a pumping scheme that involves synchronized flux and voltages signals was based on the Josephson energies modulation of two SQUIDs that forms a mesoscopic island\cite{Niskanen2005}. Observed nonidealities in the pumped charge were attributed to incomplete suppression of the Josephson energies and for this reason, a multiloop SQUID geometry was suggested. With an optimized sluice design, a pumping current of Cooper pairs in the nanoampere regime has been demonstrated\cite{Vartiainen2007}. There, the SQUIDs inhomogeneity still affected the experiment, indicating that more sophisticated topologies were required to reduce the residual Josephson energy and consequently the leakage current.
Recently, it has been theoretically and experimentally demonstrated that in nanosized superconductors the Cooper condensate can be collapsed by means of a phase difference equal to $\pi$ over a length scale comparable to the superconducting coherence length\cite{Ronzani2017}. The complete suppression of the energy gap in the local density of quasiparticle states has been probed by means of tunnel current spectroscopy in a mesoscopic fully-superconducting tunnel junction and open up several still unexplored possibilities in the field of in-situ Josephson energy modulations.
In this framework, the paper will apply a coherent pumping scheme in the non–adiabatic limit to a device proposal consisting of a superconducting sluice variant where SQUIDSs are replaced by nanosized superconductors leading to improved and simplified Josephson energy suppression.

\section{\label{sec:fulxpumping} Magnetically driven superconducting gap suppression in a two level system}

\begin{figure}
    \centering
    \includegraphics[width=\linewidth]{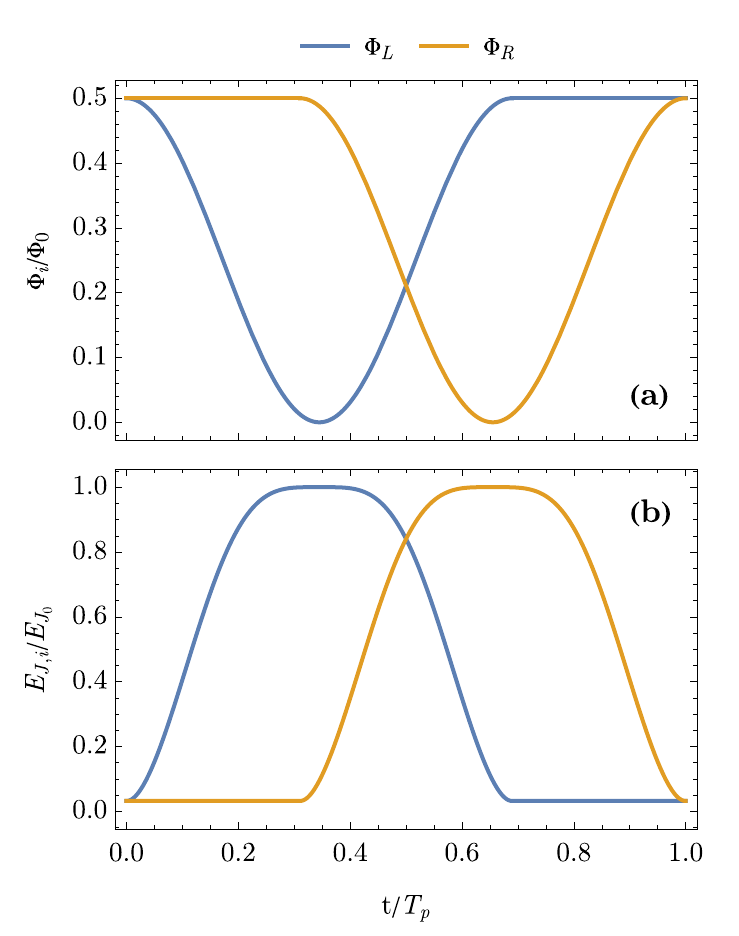}
    \caption{(a) Magnetic flux threading the left (blue) and right (orange) loops as a function of time. The modulation of the fluxes has a cosinusoidal shape with a phase lag of $1/3$ of a period. (b) Modulation of the Josephson energies caused by the flux modulation. $E_\mathrm{J_\mathrm{0}}$ is the maximum value of the Josephson energy reached in one cycle.}
    \label{fig:controlparameters}
\end{figure}

The structure of the proposed device is shown in Fig. \ref{fig:sketch}. It is composed of two Superconducting Quantum Interference Proximity Transistors\cite{giazottoSQUIPT2010} (SQUIPTs), namely superconducting loops interrupted at one side by a short gap (green rectangle) containing a thin superconducting nanowire that joins the two sides of the loop. A superconducting island (blue wire) connects the two SQUIPTs and forms at its edges two Josephson junctions with the superconducting wires. The structure made of SQUIPTs plus superconducting island recalls a Superconducting Quantum Interference Single - Electron Transistor (SQUISET) introduced in \cite{enricoSET2016,enricoSET2019}, but with the main difference that the SQUISET has a negligible Josephson energy\cite{tinkham} $E_\mathrm{J}$ compared to the charge energy of the junctions $E_\mathrm{c}$
\begin{align} 
 E_\mathrm{J} &=  \frac{\phi_0}{2\pi}I_\mathrm{c} \label{eq:josephson_energy}\\ 
 E_\mathrm{c} &=  \frac{e^2}{2C}
\end{align}

Here, $\phi_0=\frac{h}{2e}$ is the flux quantum, $h$ is the Plank constant, $e$ is the elementary charge, $I_\mathrm{c}$ and $C$ are the critical current and capacitance of the Josephson junctions. The two Josephson junctions are supposed to be identical, hence having the same $I_\mathrm{c}$ and $C$.\\
Without lack of generality, we can assume that between the two leads connecting the device to the measurement setup there is a constant and tunable bias phase difference $\phi$. Reminding the gating structure of a sluice \cite{Vartiainen2007}, a gate electrode capacitively coupled to the island sets the amount of charge $Q=C_\mathrm{G} V_\mathrm{G}$ on it through the voltage $V_\mathrm{G}$ ($C_\mathrm{G}$ is the gate capacitance), hence a number of Cooper pairs $n_\mathrm{g} = (C_\mathrm{G} V_\mathrm{G})/(2e)$.\\
Following the approach adopted in \cite{PhysRevB.82.134517,Gasparinetti2012} we switch to a different variable to quantify the charge on the island defined as $\delta n_\mathrm{g} = n_\mathrm{g} - \frac{1}{2}$, that represents the extra amount of charge on the island respect to a half elementary charge.\\
It has been demonstrated \cite{Gasparinetti2012,PhysRevLett.105.030401,PhysRevB.82.134517,pekola1999} that the pumping capabilities of such Josephson - based charge pumps can be properly described in the framework of a dynamic two levels system, where the amount of charge on the island is chosen as a quantum number. The dimension of the Hilbert space is then rescaled down to just two, corresponding to the quantum levels where there is no net charge on the island $\ket{0}$ and one cooper pair on the island $\ket{1}$. The Hamiltonian of this system can be written in matrix form as \cite{Gasparinetti2012}
\begin{equation}\nonumber
\hat{H} = 
\left(\begin{array}{cc} 
E_C(\frac{1}{2}+\delta n_\mathrm{g})^2 & E_+ \cos\frac{\phi}{2}+iE_-\sin\frac{\phi}{2}\\ E_+ \cos\frac{\phi}{2}-iE_-\sin\frac{\phi}{2} & E_C(\frac{1}{2}-\delta n_\mathrm{g})^2 \end{array}\right)
\end{equation}\label{eq:hamiltonian}

\begin{figure*}
    \centering
    \includegraphics[width=15cm]{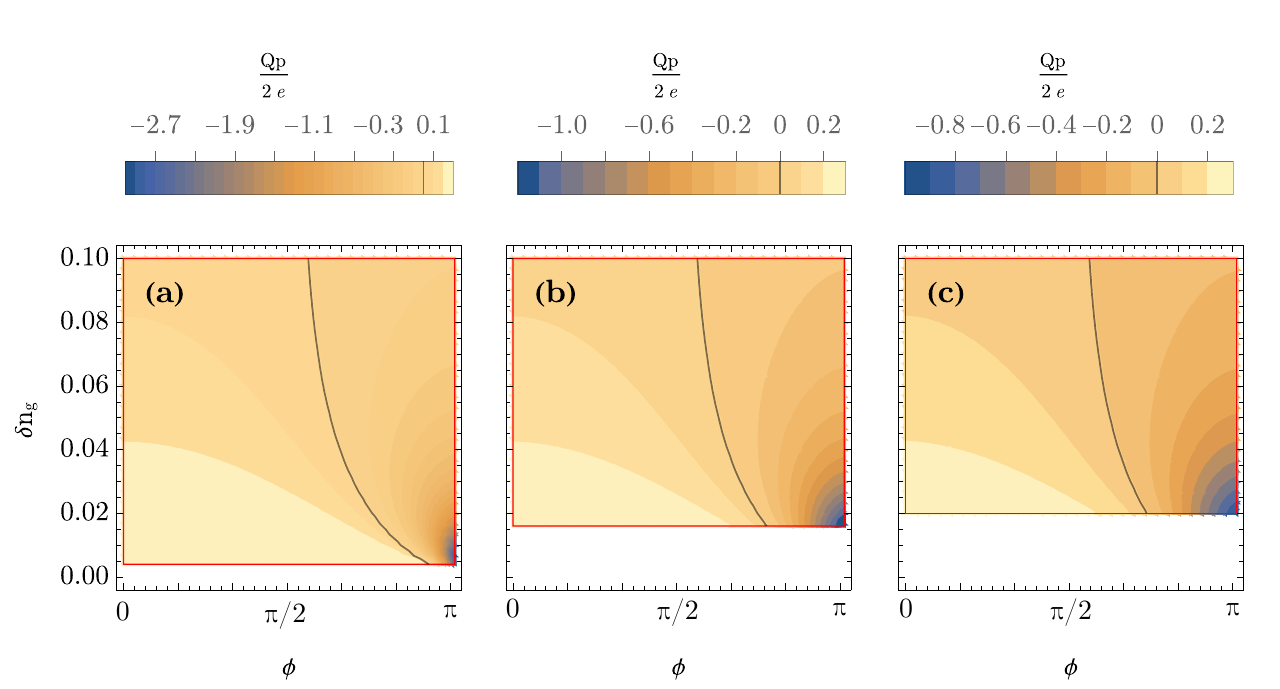}
    \caption{The plots represent the number of Cooper pairs pumped in one cycle between the two leads as a function of the phase bias $\phi$ and the gate offset $\delta n_\mathrm{g}$. Plots (a), (b) and (c) are computed respectively for a pumping frequency of 10 MHz, 100 MHz and 200 MHz. The black curve represents no net charge moved between the leads. Plots (a), (b) and (c) are plotted on ranges of $\delta n_\mathrm{g}$ which differ from each other in the lower bound, respectively 0.004, 0.016 and 0.02. This is necessary to sit in the validity range of the master equation approach $\delta n_\mathrm{g} \gg J_\mathrm{min}/E_c$.}
    \label{fig:contourplots}
\end{figure*}

where it has been defined $E_{\pm}=\frac{1}{2}(E_\mathrm{J,l}\pm E_\mathrm{J,r})$ with $E_\mathrm{J,l}$ and $E_\mathrm{J,r}$ the Josephson energies of the left and right junctions defined in \eqref{eq:josephson_energy}.\\
The modulation of the Josephson energy in a pumping cycle, which we define as a cyclic modulation of the magnetic fluxes $\Phi_\mathrm{i}$ into the loops, enables the net charge transfer from a lead to the other. This modulation can be achieved through a variation of the magnetic flux threading a superconducting loop\cite{D'Ambrosio2015,Ronzani2017} because the phase difference across the nanowire acts like a valve that breaks down the order parameter into the nanowire, resulting in a decrease of the Josephson energy. To express this Josephson energy modulation in terms of magnetic flux threading the left and right loops we need to link the critical currents of the Josephson junctions to the superconducting gaps of the nanowires. That is possible by recalling the Ambegaokar - Baratoff formula\cite{ambegaokar1963} which predicts the critical current of a weak link formed by an oxide layer
\begin{equation}\label{eq:ambegaokar-baratoff}
    I_\mathrm{c} = \frac{\Delta_1}{R_\mathrm{n}e}K\Bigg(\sqrt{1-\frac{\Delta_1}{\Delta_2}}\Bigg)
\end{equation}

where $R_\mathrm{n}$ is the normal state resistance of the junction, $K(x)$ is the complete elliptic integral of the first kind with argument $x$ and $\Delta_1$ and $\Delta_2$ are respectively the energy gaps of the superconducting leads forming a Josephson junction ($\Delta_1 < \Delta_2$), namely the nanowires and the superconducting island.\\
A change in the magnetic flux threading a loop will change accordingly the phase difference across the corresponding nanowire. For a proximized nanowire, it has been shown\cite{Ronzani2014} that a proper phase difference tuning can lead to a suppression of its superconducting density of states and so to a reduction of its superconducting energy gap $\Delta_\mathrm{i}$. This modulation can be analytically expressed as
\begin{align}\label{eq:delta_i}
    \Delta_\mathrm{i} = \Delta_\mathrm{min} + (\Delta_\mathrm{max} - \Delta_\mathrm{min})\cdot\abs{\cos{\Big(\pi\frac{\Phi_\mathrm{i}}{\phi_0}\Big)}}
\end{align}
with $\Delta_\mathrm{max}$ and $\Delta_\mathrm{min}$ respectively the minimum and maximum values of the gap reached during the modulation and $\Phi_\mathrm{i}$ the magnetic flux threading the i-th loop. Equation \eqref{eq:delta_i} gives therefore the link between the control parameter $\Phi_\mathrm{i}$ and the Josephson energy expressed by equation \eqref{eq:josephson_energy}. It is possible seeing that by joining equations \eqref{eq:josephson_energy}, \eqref{eq:ambegaokar-baratoff} and \eqref{eq:delta_i}, one can obtain modulation of the Josephson energy due to a change in the $\mathrm{i}$-th magnetic flux.\\
Under the assumption that the control parameters $\Phi_\mathrm{i}$ will slowly change with time, we can lean on a quasi-adiabatic description of the system. To do that we introduce a local adiabatic parameter $\alpha(t) = \abs{\bra{\dot{g}(t)}\ket{e(t)}}/\Delta(t)$, where $\ket{g(t)}$ and $\ket{e(t)}$ are the instantaneous ground and first excited states and $\Delta(t)$ is the instantaneous energy gap between the two levels (the dot sign indicates time derivation).
The quasi-adiabatic assumption implies that at each moment $\alpha\ll1$, ensuring that the system will remain in its ground state for the entire duration of the $\Phi_\mathrm{i}$ modulation. It is useful to express the ground state and first excited state in terms of the charge number base
\begin{align}
    \ket{g} = \frac{1}{\sqrt{2}}(\sqrt{1-\eta}\ket{0} + e^{-i\gamma}\sqrt{1+\eta}\ket{1})\\
    \ket{e} = \frac{1}{\sqrt{2}}(\sqrt{1+\eta}\ket{0} - e^{-i\gamma}\sqrt{1-\eta}\ket{1})
\end{align}

defining
\begin{align}
    \gamma =& \arctan(\frac{E_\mathrm{J,r} - E_\mathrm{J,l}}{E_\mathrm{J,r} + E_\mathrm{J,l}}\tan\frac{\phi}{2})\\
    \eta =& \frac{\delta n_\mathrm{g}}{\sqrt{\delta n_\mathrm{g}^2+(\frac{E_{12}}{E_\mathrm{C}})^2}}\\
    E_{12} =& \frac{1}{2}\sqrt{E_\mathrm{J,l}^2 + E_\mathrm{J,r}^2 + 2E_\mathrm{J,l}E_\mathrm{J,r}\cos{\phi}}
\end{align}

In the following the master equation approach followed by \cite{solinas2010,Gasparinetti2012,PhysRevLett.105.030401} is adopted to calculate the amount of pumped charge as a function of the experimental parameters, with particular attention to the frequency dependence. This approach is useful when dealing with decoherence caused by the interaction between the two-level system and the environment and allows to explore its quasi-adiabatic dynamics. Non-adiabatic transitions can occur due to the finite frequency at which the pumping cycle is performed, hence where the adiabatic approximation loses validity. Quasi-adiabaticity and decoherence are taken into account by this approach through the density matrix elements $\rho_{\mathrm{gg}}$ and $\rho_{\mathrm{ge}}$, where the subscripts g and e correspond respectively to the ground and excited states. In this framework, the effect of the interaction with the environment is given by a fictitious resistor $R$ capacitively coupled to the island \cite{PhysRevB.82.134517}. Referring to the convention used in \cite{PhysRevB.82.134517} we take $R=300$ k$\Omega$ and $g=0.02$ as its coupling constant. The temperature of the resistor represents the temperature of the environment, which will be kept zero during all the calculations \cite{Gasparinetti2012}. In this way, we make use of the following expression for the observables\cite{PhysRevB.82.134517}

\begin{align}
    Q_\mathrm{d,l} =& \frac{2e}{\hbar}\int_{0}^{T_\mathrm{p}}\frac{\rho_\mathrm{gg}E_\mathrm{J,l}\sqrt{1-\eta^2}}{2}\sin{(\eta+\frac{\phi}{2})}dt\\
     Q_\mathrm{d,r} =& -\frac{2e}{\hbar}\int_{0}^{T_\mathrm{p}}\frac{\rho_\mathrm{gg}E_\mathrm{J,r}\sqrt{1-\eta^2}}{2}\sin{(\eta-\frac{\phi}{2})}dt\\
    Q_\mathrm{p,l} =& \frac{2e}{\hbar}\int_{0}^{T_\mathrm{p}}E_\mathrm{J,l}
    [\eta \mathrm{Re}(\rho_\mathrm{ge})\sin{(\eta+\frac{\phi}{2})}-\\
    &-\cos{(\eta+\frac{\phi}{2})}\mathrm{Im}(\rho_\mathrm{ge})]dt\\
    Q_\mathrm{p,r} =& \frac{2e}{\hbar}\int_{0}^{T_\mathrm{p}}E_\mathrm{J,r}
    [\cos{(\eta-\frac{\phi}{2})}\mathrm{Im}(\rho_\mathrm{ge})-\\
    &-\eta \mathrm{Re}(\rho_\mathrm{ge})\sin{(\eta-\frac{\phi}{2})}]dt
\end{align}

where $Q_\mathrm{d}$ and $Q_\mathrm{p}$ are respectively the dynamic and pumped charge moved between the two leads in one cycle of period $T_\mathrm{p}$. $Q_\mathrm{p}$ is the charge moved from one side of the device to the other solely due to the modulation of the magnetic fluxes, hence the control parameters. On the contrary $Q_\mathrm{d}$ incorporates the quantum phase coherence effects that follow from a phase gradient into a superconductor.

\subsection{\label{sec:results}Results and discussion}

\begin{figure}
\includegraphics[width=0.5\textwidth]{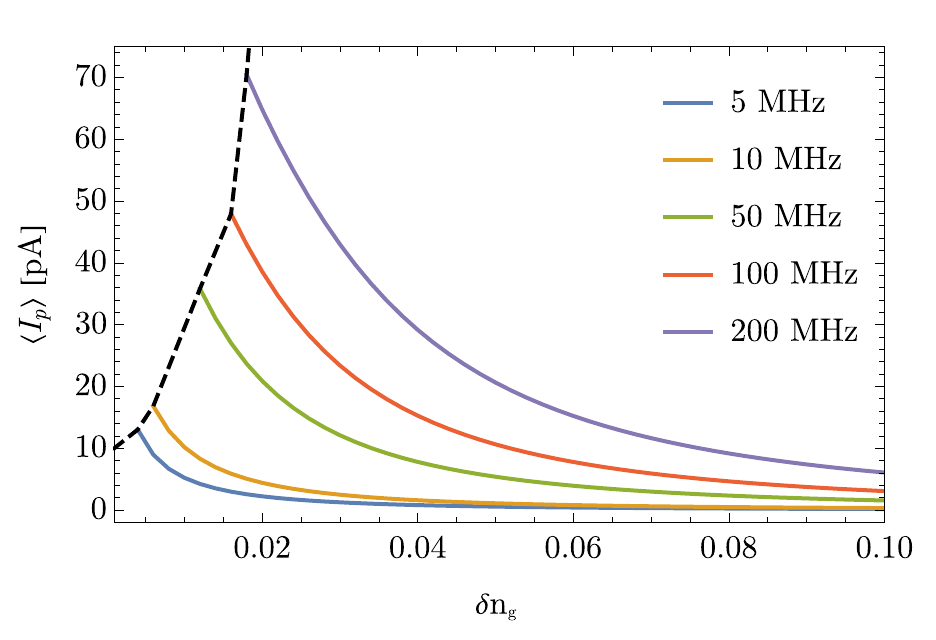}
\caption{\label{fig:averagecurrent} Average pumped current calculated using Equation \eqref{eq:pumped_current} as a function of $\delta n_\mathrm{g}$ for different frequencies of the pumping cycle. The curves are terminated for values of $\delta n_\mathrm{g}$ that bring the model out of its validity range ($\delta n_\mathrm{g} \gg J_\mathrm{min}/E_\mathrm{c}$). The black dashed curve is used to guide the eye on this validity limit.}
\end{figure}

\begin{figure}
\includegraphics[width=0.5\textwidth]{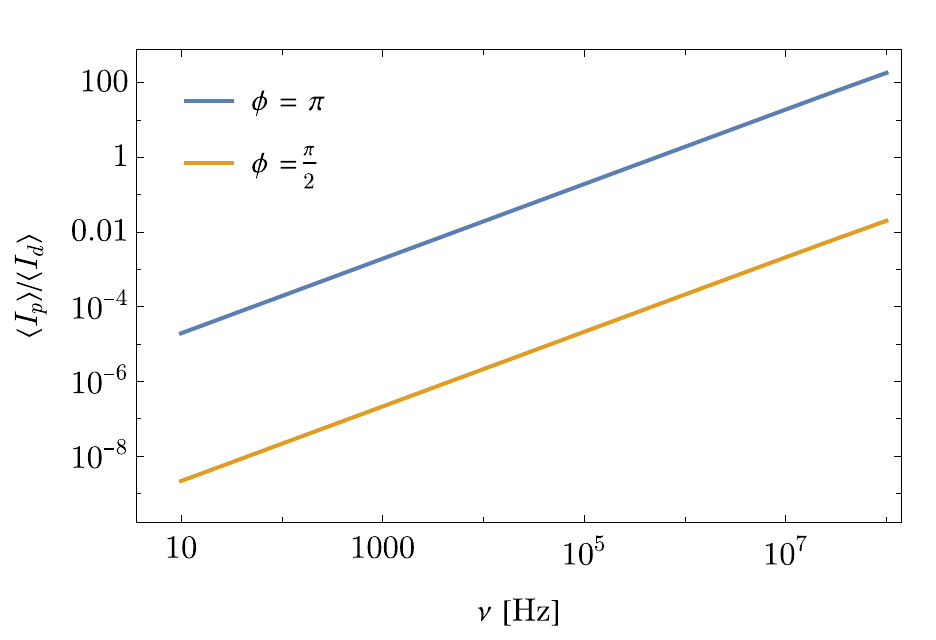}
\caption{\label{fig:currentratio} Pumped current on dynamic current ratio $\langle I_\mathrm{p}\rangle/\langle I_\mathrm{d}\rangle$. The curves are calculated as a function of the pumping frequency $\nu$ for different bias phases and $\delta n_\mathrm{g} = 0.03$.}
\end{figure}

Commonly used BCS superconductors for the production of Single-Electron Transistors and Cooper pair pumps are Al and Nb, the former for the superior quality of its oxides in the fabrication of the weak links while the latter for its high superconducting gap. For this reason, we choose Al for the two loops and the nanowires (critical temperature of 1.2 K) while we choose Nb for the island (critical temperature 9.3 K). The normal state resistance of the Josephson junctions is taken as $170 \; \mathrm{k}\Omega$ for both junctions while their charge energy is $E_\mathrm{c}=1\:\mathrm{K}\cdot k_\mathrm{B}$, where $k_\mathrm{B}$ is the Boltzman constant.\\
As mentioned above, the pumping capabilities of the device can be estimated considering a particular pumping cycle for the control parameters. The chosen modulation of the control parameters are presented in Fig. \ref{fig:controlparameters} (a). The magnetic flux has a cosine like modulation that brings the total flux threading a loop from a half flux quantum down to $15\% \Delta_\mathrm{max}$ in a cycle (hence $\Delta_\mathrm{min}=0.15\cdot \Delta_\mathrm{max}$ in equation \eqref{eq:delta_i}). A cosine like modulation has been chosen to reduce the needed bandwidth in a hypothetical practical realization so that a more realistic pumping cycle could be modelled. $\Phi_\mathrm{L}$ and $\Phi_\mathrm{R}$ are modulated with a time shift of $T_\mathrm{P}/3$. This shift has been chosen through a first optimization process of the pumped charge as a function of the phase shift between the two flux waves, then the shift was kept constant for all the following computations. The resultant Josephson energy modulation is shown in Fig. \ref{fig:controlparameters} (b).\\
The average pumped charge was calculated exploring the parameter space in terms of the bias phase difference $\phi$ and the gate offset $\delta n_\mathrm{g}$. Fig. \ref{fig:contourplots} shows some contour plots of the average pumped charge in one cycle, evaluated for different pumping frequencies. The black contour line corresponds to the zero pumped charge points, hence the condition in the $\left(\phi, \delta n_\mathrm{g}\right)$ plane where no net charge is moved between the leads nonetheless a pumping cycle is being performed. In order to be sure to sit in the validity range of the master equation where non-adiabatic transitions are not the dominant drive of the system, a threshold condition on the amplitude of the matrix element $\rho_\mathrm{ge}$ needs to be fulfilled. The validity of the adiabatic theorem requires the value of the matrix element $\rho_\mathrm{ge}$ to remains beyond unity, so setting the arbitrary threshold $\rho_\mathrm{ge}<0.3$ (see reference\cite{Gasparinetti2012} for further details) we are sure not to go beyond this limits of validity. This threshold can be indirectly seen in plots (a), (b) and (c) of Fig. \ref{fig:contourplots}, where the coordinate $\delta n_{\mathrm{g}}$ extends differently for the three plots, with values corresponding to
\begin{equation}\label{eq:delta_ng}
    \delta n_\mathrm{g} \gg J_\mathrm{min}/E_\mathrm{c}
\end{equation}

with $J_\mathrm{min}$ the lowest value of the Josephson energy in a pumping cycle.
From Fig. \ref{fig:contourplots} it is clear that the higher the pumping frequency the lower is the pumped charge per cycle. However, this is not surprising since it is known \cite{Gasparinetti2012} that more charge is moved when pumping slowly, indeed in the adiabatic limit, hence where an infinitely slow pumping cycle is performed, electron and Cooper pair pumps show the best performance. Nonetheless, the measured quantity in experiments is the average pumped current
\begin{equation}\label{eq:pumped_current}
   \langle I_\mathrm{p} \rangle = \langle Q_\mathrm{p}\rangle\cdot\nu
\end{equation}
with $\nu=1/T_\mathrm{p}$ is the pumping frequency. Observing Fig. \ref{fig:contourplots} we notice that the maximum pumped charge per cycle can be always obtained for a phase bias close to $\pi$. This fact suggests that to obtain the maximum possible pumped current we should keep $\phi = \pi$ as the phase difference across the device. In Fig. \ref{fig:averagecurrent} we report the average pumped current expressed in Equation \eqref{eq:pumped_current} for different pumping frequencies as a function of $\delta n_\mathrm{g}$. From this plot we can infer that nonetheless the pumped charge decreases for higher frequencies, increasing the number of cycles per second it is possible to compensate for this effect by obtaining a global rise of the pumped current. The curves are ended for different values of $\delta n_\mathrm{g}$ to take into account condition \eqref{eq:delta_ng}.\\
A figure of merit of any electron/Cooper pair pump is represented by the pumped current on dynamic current ratio $\langle I_\mathrm{p}\rangle/\langle I_\mathrm{d}\rangle$, where $\langle I_\mathrm{d} \rangle = \langle Q_\mathrm{d}\rangle\cdot\nu$. The dynamic current represents that component of the current that is independent of the dynamics of the control parameters and that depends only on the phase difference across the device. If we consider a pumping cycle (Fig. \ref{fig:controlparameters}) it can be shown that the dynamic current is non-zero just when both the Josephson energies are non-zero at the same time. This fact is not surprising if we remember that the whole device is superconducting and we bias it with a constant phase difference, indeed when we suppress the Cooper condensate into the nanowires bringing the Josephson energies to zero, the device becomes locally non-superconducting, meaning that the supercurrent cannot flow due to the sole phase difference $\phi$ between the two leads. For this reason, the dynamic current, which can flow just in the centre of a pumping period where both $E_\mathrm{J,L}$ and $E_\mathrm{J,R}$ are non-zero, has the footprint of a leakage current. It is thus interesting studying the $\langle I_\mathrm{p}\rangle/\langle I_\mathrm{d}\rangle$ ratio to quantify the relative contribution of these two components with respect to the total current flowing between the two leads. In Fig. \ref{fig:currentratio} the $\langle I_\mathrm{p}\rangle/\langle I_\mathrm{d}\rangle$ ratio as a function of the pumping frequency for two values of the phase bias is shown. The pumped current $\langle I_\mathrm{p}\rangle$ (see Fig. \ref{fig:averagecurrent}) remains of the same order of magnitude in all the explored pumping frequency range, on the contrary, the dynamic current decreases linearly increasing the pumping frequency. This is due to the fact that the time window during which $I_\mathrm{d}$ can flow from one lead to the other gets shorter and shorter with the growing frequency. The net result is that for high enough frequencies $I_\mathrm{p}$ becomes dominant on $I_\mathrm{d}$, making $I_\mathrm{d}$ just a small offset on the total current. As already mentioned, the pumped current follows the direction of the pumping cycle, on the contrary, the dynamic current goes in the direction of the gradient of the bias phase difference. This fact is of key importance in all the experimental realizations since it allows to discern between the two components just by inverting the pumping direction and then subtracting the offset out, giving the net pumped current \cite{Mottonen2008}. 

\subsection{\label{sec:conclusions}Conclusions}

A novel concept of a fully superconducting Cooper pair pump has been presented. The ability to pump charge coherently emerges thanks to the cyclic modulation of the magnetic flux threading the two superconducting loops composing the leads of the device, without the need of a modulation of the gate voltage, which on the contrary determines, together with the overall phase bias, the working point. The study of the pumped charge as a function of the phase bias and the gate offset underline how the device performs better set in a working point with the bias phase difference $\phi=\pi$, reaching a maximum amount of $\approx3$ Cooper pairs moved per cycle at low frequencies. The study of the average pumped current as a function of the pumping frequency highlights the effect of decoherence and non-adiabatic transitions through the lowering of the pumped charge per cycle at high frequencies, that for the higher frequencies becomes a fraction of a Cooper pair. An important figure of merit in practical realizations is represented by the amount of pumped current compared to the dynamic current. Studying the $\langle I_\mathrm{p}\rangle/\langle I_\mathrm{d}\rangle$ ratio we find out that the higher the pumping frequency is and the less predominant becomes the dynamic component on the total current.

\end{document}